%% file: main.tex
\documentclass[a4paper,11pt]{article}
\usepackage{pos}
 \usepackage[nice]{nicefrac}
\def\HP{\hphantom{\alpha}} % horizontal
\usepackage[utf8]{inputenc}
\usepackage{graphicx}
\usepackage{amsmath}
\usepackage{amsfonts}
\usepackage{physics}
\usepackage{xspace}
\usepackage{bm}
\usepackage{mathrsfs}
\usepackage{nicefrac}
\usepackage{multirow}
\usepackage[format=plain,justification=RaggedRight,labelsep=endash,font=small,labelfont=sc]{caption}
\usepackage{stackengine}

\usepackage[export]{adjustbox}
\usepackage{comment}

\usepackage{yfonts}

\def\GLW{{\rm GLW}}

\definecolor {darkgreen}{rgb}{0.2,0.7,0.2}

\def\be{\begin{equation}}
	\def\ee{\end{equation}}
\newcommand{\bel}[1]{\begin{eqnarray}\label{#1}}
	\newcommand{\eel}{\end{eqnarray}}
\def\barr{\begin{array}}
	\def\earr{\end{array}}
\def\beq{\begin{eqnarray}}
	\def\eeq{\end{eqnarray}}
\def\bfig{\begin{figure}}
	\def\efig{\end{figure}}
\def\lt{\left}
\def\rt{\right}
\def\CHI{\chi}
\newcommand{\nn}{\nonumber}
\def\spin{\,\textgoth{s:}}

\def\a{\alpha}
\def\b{\beta}
\def\g{\gamma}
\def\d{\delta} 
\def\r{\rho}
\def\s{\sigma}
\def\c{\chi}
 
\def\LR{\left(} % round
\def\RR{\right)}
\def\LS{\left[} % square
\def\RS{\right]}
\def\LC{\left{} % curly
\def\RC{\right}}
\def\LA{\left\langle}% angle
\def\RA{\right\rangle}
\def\LD{\left.}% dummy
\def\RD{\right.}
\def\HP{\hphantom{\alpha}} % horizontal

\def\half{\frac{1}{2}}

\def\GLW{{\rm GLW}}
\def\LRFF{{\rm LRFF}}

\def\nU{n_{(0)}}
\def\eU{\varepsilon_{(0)}}
\def\PU{P_{(0)}}
\def\sU{s_{(0)}}

\def\nP{n_{}}
\def\eP{\varepsilon_{}}
\def\PP{P_{}}
\def\sP{s_{}}
\def\wP{w_{}}

\def\pmu{p^\mu}
\def\pnu{p^\nu}

\def\vv{{\boldsymbol v}}
\def\pv{{\boldsymbol p}}
\def\av{{\boldsymbol a}}
\def\bv{{\boldsymbol b}}
\def\kv{{\boldsymbol k}}
\def\omnL{\omega_{\mu\nu}}
\def\omnU{\omega^{\mu\nu}}
\def\omnLbar{{\bar \omega}_{\mu\nu}}
\def\omnUbar{{\bar \omega}^{\mu\nu}}
\def\omnLbardot{{\dot {\bar \omega}}_{\mu\nu}}
\def\omnUbardot{{\dot {\bar \omega}}^{\mu\nu}}

\def\oabL{\omega_{\alpha\beta}}
\def\oabU{\omega^{\alpha\beta}}
\def\omnLD{{\tilde \omega}_{\mu\nu}}
\def\omnUD{\tilde {\omega}^{\mu\nu}}
\def\omnLDbar{{\bar {\tilde \omega}}_{\mu\nu}}
\def\omnUDbar{{\bar {\tilde {\omega}}}^{\mu\nu}}

\def\epsLmnbg{\epsilon_{\mu\nu\beta\gamma}}
\def\epsUmnbg{\epsilon^{\mu\nu\beta\gamma}}
\def\epsLmnab{\epsilon_{\mu\nu\alpha\beta}}
\def\epsUmnab{\epsilon^{\mu\nu\alpha\beta}}

\def\epsUmnrs{\epsilon^{\mu\nu\rho \sigma}}
\def\epsUlnrs{\epsilon^{\lambda \nu\rho \sigma}}
\def\epsUlmrs{\epsilon^{\lambda \mu\rho \sigma}}

\def\epsLmnbg{\epsilon_{\mu\nu\beta\gamma}}
\def\epsUmnbg{\epsilon^{\mu\nu\beta\gamma}}
\def\epsLmnab{\epsilon_{\mu\nu\alpha\beta}}
\def\epsUmnab{\epsilon^{\mu\nu\alpha\beta}}

\def\epsLabgd{\epsilon_{\alpha\beta\gamma\delta}}
\def\epsUabgd{\epsilon^{\alpha\beta\gamma\delta}}

\def\epsUmnrs{\epsilon^{\mu\nu\rho \sigma}}
\def\epsUlnrs{\epsilon^{\lambda \nu\rho \sigma}}
\def\epsUlmrs{\epsilon^{\lambda \mu\rho \sigma}}

\def\epsLijk{\epsilon_{ijk}}
\def\half{\frac{1}{2}}

\def\GLW{{\rm GLW}}

\def\n0{n_{(0)}}
\def\e0{\varepsilon_{(0)}}
\def\P0{P_{(0)}}
\title{Relativistic magnetohydrodynamics with spin}
\author[a]{Samapan Bhadury} 
\affiliation[a]{Institute of Theoretical Physics, Jagiellonian University, ul. St. \L ojasiewicza 11, 30-348 Krakow, Poland}
\emailAdd{bhadury.samapan@gmail.com}
\author[a]{Wojciech Florkowski}
\emailAdd{wojciech.florkowski@uj.edu.pl}
\author[b]{Amaresh Jaiswal}
\affiliation[b]{School of Physical Sciences, National Institute of Science Education and Research, An OCC of Homi Bhabha National Institute, Jatni-752050, India}
\emailAdd{a.jaiswal@niser.ac.in} 
\author[b,c]{Avdhesh~Kumar} 
\affiliation[c]{Institute of Physics, Academia Sinica, Taipei, 11529, Taiwan}
\emailAdd{avdheshk@gate.sinica.edu.tw}
\author*[d]{Radoslaw Ryblewski} 
\affiliation[d]{Institute of Nuclear Physics Polish Academy of Sciences, PL-31-342 Krakow, Poland}
\emailAdd{radoslaw.ryblewski@ifj.edu.pl}

\abstract{
In this work, we present a novel framework of relativistic non-resistive dissipative magnetohydrodynamics for spin-polarized particles. Utilizing a classical relativistic kinetic equation for the distribution function in an extended phase-space of position, momentum, and spin, we derive equations of motion for dissipative currents at first-order in spacetime gradients. Our findings reveal a coupling between fluid vorticity and magnetization via an electromagnetic field, leading to relativistic analogs of the Einstein-de Haas and Barnett effects. Our study provides a tool for a better understanding of the polarization phenomena observed in relativistic heavy-ion collisions.
}

\FullConference{%
  25th International Spin Symposium (SPIN 2023)\\
  24-29 September, 2023}

\begin{document}
\include{commands}
\maketitle
\section{Introduction}
\label{sec:introduction}
Over the last decades, it has been well established~\cite{Gale:2013da} that the strongly interacting matter produced in relativistic nuclear collisions evolves according to principles of relativistic hydrodynamics~\cite{Florkowski:2017olj,Rocha:2023ilf}.  
It is expected that in non-central collisions this matter may experience large angular momentum and a strong magnetic field~\cite{Tuchin:2013ie,Becattini:2007sr}.
These extreme physical conditions may lead, similarly to the non-relativistic magneto-mechanical effects of Einstein-de Haas \cite{Einstein} and Barnett \cite{Barnett}, to spin polarization and magnetization of the matter and, consequently, of the emitted particles \cite{Liang:2004ph, %Liang:2004xn, 
Voloshin:2004ha,Betz:2007kg}. 
The existence of spin polarization phenomenon was recently confirmed experimentally~\cite{%STAR:2018gyt,
STAR:2017ckg,STAR:2019erd,ALICE:2019aid,ALICE:2019onw, STAR:2020xbm, STAR:2021beb, HADES:2022enx}
triggering vast theoretical developments aiming at finding a unified interpretation of the measured observables~\cite{Becattini:2007nd, Becattini:2009wh,Becattini:2013fla,Li:2017slc,Sun:2017xhx,Becattini:2018duy, Florkowski:2018ahw,Wu:2019eyi,Sheng:2019kmk,Fu:2020oxj,Yang:2020hri,Deng:2020ygd, Ambrus:2020oiw,Palermo:2021hlf,Florkowski:2021pkp,Li:2021zwq,Yi:2021ryh,Kumar:2022ylt,Kumar:2023ghs}.
In particular, based on fundamental conservation laws, an extension of relativistic hydrodynamics for spin-polarized fluids was proposed~\cite{Florkowski:2017ruc} giving rise to the rapid development of a new field known as relativistic spin hydrodynamics~\cite{Florkowski:2017dyn,Florkowski:2018fap,Hattori:2019lfp,Speranza:2020ilk,Bhadury:2020cop,Bhadury:2020puc,Hu:2021pwh, 
Shi:2020htn,Weickgenannt:2020aaf,Speranza:2021bxf,She:2021lhe,Peng:2021ago,Wang:2021ngp,Yi:2021unq,Florkowski:2019qdp,Singh:2021man,Florkowski:2021wvk,Montenegro:2017rbu,Gallegos:2021bzp,Hongo:2021ona,Yi:2021unq,Gallegos:2022jow,Ambrus:2022yzz,Weickgenannt:2022zxs,Daher:2022xon,Daher:2022wzf,Biswas:2022bht,Weickgenannt:2022qvh,Dey:2023hft,Biswas:2023qsw,Bhadury:2023vjx,Weickgenannt:2023btk,Becattini:2023ouz,Kiamari:2023fbe,Daher:2024ixz}.

Very recently, a formalism of dissipative non-resistive spin magnetohydrodynamics was constructed, aiming at incorporating into the spin hydrodynamics effects of spin polarization due to the presence of electromagnetic field~\cite{Bhadury:2022ulr}. In this contribution, we briefly review the framework of~\cite{Bhadury:2022ulr} and discuss its main implications. Starting from the classical transport equation for the distribution function in an extended phase-space of position, momentum, and spin, in the presence of a magnetic field we derive equations of motion for dissipative currents at first-order in spacetime gradients. It is found that, apart from contributions from various standard hydrodynamic gradients~\cite{Bhadury:2020puc, Bhadury:2020cop}, the spin current acquires also effects due to the gradients of electromagnetic field~\cite{Bhadury:2022ulr}. In particular, we show that the coupling between fluid vorticity and magnetization via an electromagnetic field gives rise to effects similar to that of Einstein-de Haas and Barnett. 

We use the following conventions for the metric tensor and Levi-Civita symbol: $g_{\mu\nu} =  \hbox{diag}(+1,-1,-1,-1)$ and $\epsilon^{0123} = -\epsilon_{0123}=1$. We also use natural units with $c = \hbar = k_B =1$.
%
%%%%%%%%%%%%%%%%%%%%%%%%%%%%%%%% 
\section{Kinetic theory derivation of equations of motion}
\label{sec:Kin_th}
%%%%%%%%%%%%%%%%%%%%%%%%%%%%%%%%
%
We consider the classical distribution function of particles with spin in an extended phase-space of space-time position $x\equiv x^\mu$, four-momentum $p\equiv p^\mu$, and intrinsic angular momentum $s\equiv s^{\mu\nu}$, $f\equiv f(x,p,s)$ \cite{Florkowski:2018fap}. The dynamics of $f$ is determined by the following kinetic equation~\cite{Bhadury:2022ulr}
\begin{align}
    %1
    \left( p^\alpha \dfrac{\partial }{\partial x^\alpha} + m\,\mathcal{F}^\alpha \dfrac{\partial }{\partial p^\alpha} + m\,\mathcal{S}^{\alpha\beta} \dfrac{\partial }{\partial s^{\alpha\beta}} \right) f &= C [\,f\,]\,,\label{NCL:Beq_v1}
\end{align}
and likewise for anti-particles with the replacement $f\to\bar{f}$. In Eq.~(\ref{NCL:Beq_v1}), the four-momentum $p^\mu = (E_p, \boldsymbol{p})$ is on the mass shell, with $E_p = \sqrt{m^2 + \boldsymbol{p}^2}$ difining the particle energy and $m$ denoting the particle mass, and $C [f]$ is the collision kernel. 

In the above equation, $\mathcal{F}^\alpha=dp^\alpha/d\tau$  and $\mathcal{S}^{\alpha\beta}=ds^{\alpha\beta}/d\tau$ (where $\tau$ denotes the proper time along the world line) are, respectively, force and torque experienced by a particle moving under influence of electromagnetic field. For composite particles they have the forms
\begin{eqnarray}
    %1
    \mathcal{F}^\alpha &=& \frac{\mathfrak{q}}{m}\,  F^{\alpha\beta} p_\beta + \frac{1}{2} \left( \partial^\alpha F^{\beta\gamma} \right) m_{\beta\gamma}, \label{NCL:Beq_Force}\\
    %2
    \mathcal{S}^{\alpha\beta} &=& 2\, F^{\gamma[\alpha}\, m_{~~\gamma}^{\beta]} - \frac{2}{m^2} \!\left(\! \chi - \frac{\mathfrak{q}}{m} \!\right)\! F_{\phi\gamma}\, s^{\phi[\alpha}\, p^{\beta]} p^\gamma , \label{NCL:Beq_Spin}
\end{eqnarray}
where $F^{\mu\nu}$ denotes the electromagnetic field strength tensor and $m^{\alpha\beta}=\chi s^{\alpha\beta}$ is the magnetic dipole moment of particles with $\chi$ playing the role of the gyromagnetic ratio \cite{Weickgenannt:2019dks}. The expressions for the first and second term on the right-hand side of Eq.~\eqref{NCL:Beq_Force} represent well-known Lorentz and Mathisson force, respectively \cite{Weickgenannt:2019dks}. On the other hand, the form of the torque in Eq.~\eqref{NCL:Beq_Spin} is less understood. Hence, in this work, we choose to neglect it.  

The number current $N^\lambda$, the energy-momentum tensor $T^{\lambda\mu}_{\rm f}$, and the spin current $S^{\lambda,\mu\nu}$ of the fluid are expressed, respectively, through the zeroth, first, and ``spin''  moment  of the distribution function~\cite{Bhadury:2020puc}
\begin{eqnarray}
N^\lambda &=& \int_{p,s} \, p^\lambda \left(f - \bar{f} \right), \label{eq:classNTS1} \\
T^{\lambda\mu}_{\rm f} &=& \int_{p,s} \, p^\lambda p^\mu \left(f + \bar{f} \right), \label{eq:classNTS2} \\
S^{\lambda,\mu\nu} &=& \int_{p,s} \, p^\lambda s^{\mu\nu} \left(f + \bar{f} \right),
\label{eq:classNTS3}
\end{eqnarray}
while the polarization-magnetization tensor is given by the formula
\begin{equation}\label{mag-pol-kin}
M^{\mu\nu} = m\int_{p,s} \, m^{\mu\nu} \left(f - \bar{f} \right).
\end{equation}
In the above equations we used the shorthand notation $\int_{p,s}\equiv \int dP dS$ with $dP \equiv  d^3p/[E_p (2 \pi )^3]$ and $dS \equiv m/(\pi \spin) \,  d^4s \, \delta(s \cdot s + \spin^2) \, \delta(p \cdot s)$, where the length of the spin vector, $\spin^2 = \frac{1}{2} \left( 1+ \frac{1}{2}  \right) = \frac{3}{4}$, is defined by the eigenvalue of the Casimir operator. 

Presuming that the microscopic interactions preserve fundamental conservation laws the following moments of the collision kernel should vanish:
\begin{equation}
    \int_{p,s} C [f] = 0, \qquad \int_{p,s} p^\mu C [f]=0, \qquad \int_{p,s} s^{\mu\nu}C[f]=0.
    \label{match}
\end{equation}
Using these properties and Eqs.~\eqref{eq:classNTS1}-\eqref{mag-pol-kin} one may show that the zeroth, first and ``spin'' moment of the kinetic equation~\eqref{NCL:Beq_v1} (assuming no torque) lead, respectively, to the following equations
\begin{equation}
\label{N^mu_cons}
    \partial_\mu N^\mu = 0, \qquad
    \partial_\nu T^{\mu\nu}_{\mathrm{f}} = F^{\mu}_{~\,\alpha} J^\alpha_{\mathrm{f}} + \frac{1}{2} \left( \partial^\mu F^{\nu\alpha} \right) M_{\nu\alpha}, 
    \qquad
    \partial_\lambda S^{\lambda,\mu\nu} = 0,
\end{equation}
where $J^\mu_{\rm f}=\mathfrak{q}N^\mu$ is a charge current with $\mathfrak{q}$ denoting the electric charge of the particles. Equations~\eqref{N^mu_cons} constitute the basis for the framework of spin-magnetohydrodynamics.
 
Assuming Landau's definition of four-velocity $u$ of the fluid, $T^{\mu\nu}_{\mathrm f}u_\nu=\epsilon u^\mu$, where $\epsilon$ is the energy density, the particle current, and the stress-energy tensor are given by
\begin{equation}\label{Nmu}
    N^\mu = n u^\mu + n^\mu, \quad
    T^{\mu\nu}_{\mathrm{f}} = \epsilon u^\mu u^\nu - \left(P + \Pi\right) \Delta^{\mu\nu} + \pi^{\mu\nu} %\label{T^mn_f},
\end{equation}
where $n$ is the net particle number density, $n^\mu$ particle number diffusion, $P$ is the pressure, $\Pi$ and $\pi^{\mu\nu}$ are the bulk and shear viscous pressures, and $\Delta^{\mu\nu} = g^{\mu\nu} - u^\mu u^\nu$.  Since we are interested in the formulation of magnetohydrodynamics with spin in the non-resistive limit, we have
\begin{equation}
F^{\mu\nu}=\epsilon^{\mu\nu\alpha\beta}u_\alpha B_\beta, \label{Fnonres}
\end{equation}
where $B^\mu$ is the magnetic field four-vector satisfying the well-known Maxwell equations, see Ref.~\cite{Bhadury:2022ulr}. The field strength tensor and polarization-magnetization tensors are related to each other by $H^{\mu\nu} - M^{\mu\nu}= F^{\mu\nu} 
$
%\footnote{Note the mostly-minus metric convention.}
, where $H^{\mu\nu}$ is the induction tensor.
%
%%%%%%%%%%%%%%%%%%%%%%%%%%%%%%%%
\section{Dynamics of dissipative currents}
\label{sec:diss_hyd}
%%%%%%%%%%%%%%%%%%%%%%%%%%%%%%%%
%
To derive constitutive relations for dissipative quantities in Eqs.~(\ref{Nmu}), we consider the kinetic equation (\ref{NCL:Beq_v1}), with the collision term treated in relaxation-time approximation (RTA) \cite{anderson1974relativistic}
\begin{equation}
    %1
    \left( p^\alpha \dfrac{\partial }{\partial x^\alpha} + m\,\mathcal{F}^\alpha \dfrac{\partial }{\partial p^\alpha} \right) f = -\left(u\cdot p\right)\frac{f-f_{\rm eq}}{\tau_{\rm eq}}\equiv -\left(u\cdot p\right)\frac{\delta f}{\tau_{\rm eq}},\label{NCL:Beq_RTA}
\end{equation}
where $f_{\rm eq}$ is the equilibrium distribution function and relaxation time $\tau_{\rm eq}$ is assumed to be independent of particle momentum and energy. Note that, within the RTA, the zeroth and first moments (see, respectively, the first and second equation in (\ref{match})) of the right-hand side of Eq.~(\ref{NCL:Beq_RTA}) vanish when Landau frame and matching conditions are used. Moreover, imposing the matching condition \cite{Bhadury:2020puc}
 \begin{equation}
   u_\lambda \delta S^{\lambda,\mu\nu} \equiv u_\lambda\left( S^{\lambda,\mu\nu} - S^{\lambda,\mu\nu}_{\rm eq} \right) = 0, \label{eq:LS}  
 \end{equation}
where $\delta S^{\lambda,\mu\nu}$ is the dissipative part of the spin current, also the spin moment (see the third equation in (\ref{match})) vanishes. 

Herein, we assume the equilibrium distribution to have the Fermi-Dirac form,
\begin{equation}\label{feq_FD}
    f_{\rm eq} = \left\{1+\exp\left[\beta (u\!\cdot\! p) - \xi - \frac{1}{2}\,\omega_{\mu\nu}s^{\mu\nu} \right]\right\}^{-1},
\end{equation}
and similarly for anti-particles with $\xi\to-\xi$, where $\xi\equiv \mu\beta$ and $\beta\equiv 1/T$. Here, $\omega_{\mu\nu}$ plays the role of Lagrange multiplier corresponding to spin conservation \cite{Florkowski:2017ruc} and is related to spin polarization observable via Pauli-Lubanski four-vector \cite{Florkowski:2017dyn, Florkowski:2018fap}. Considering the limit of small polarization, we can keep only terms up to linear in $\omega^{\mu\nu}$ and write
\begin{equation} \label{feq_f0}
    f_{\rm eq} = f_0 + \frac{1}{2}\omega_{\mu\nu}s^{\mu\nu}  f_0  (1-f_0),
\end{equation}
where $f_0\equiv\left\{ 1+\exp\left[\beta(u\!\cdot\!p) - \xi\right] \right\}^{-1}$.

The dissipative quantities defined in Eqs.~\eqref{Nmu} and~\eqref{eq:LS} are given in terms of the non-equilibrium corrections to the distribution function,
\begin{eqnarray}
    n^\mu &=&  \int_{p,s} \, p^{\langle \mu\rangle} \left(\delta f - \delta\bar{f} \right), \label{diss1} \qquad \quad \quad \quad
    \Pi =  \int_{p,s} \left(-\frac{1}{3}\right) \, p^{\langle \mu\rangle} p_{\langle \mu\rangle} \left(\delta f + \delta\bar{f} \right), \label{diss2} \\
    \pi^{\mu\nu} &=&   \int_{p,s} \, p^{\langle\mu} p^{\nu\rangle} \left(\delta f + \delta\bar{f} \right), \label{diss3} \qquad
    \delta S^{\lambda,\mu\nu} = \int_{p,s} \, p^\lambda s^{\mu\nu} \left(\delta f + \delta\bar{f} \right), \label{diss4}
\end{eqnarray}
where used the notation $X^{\langle\mu\rangle}\equiv\Delta^{\mu}_{\alpha}X^\alpha$ and $X^{\langle\mu\nu\rangle}\equiv\Delta^{\mu\nu}_{\alpha\beta}X^{\alpha\beta}$. 

To obtain the relativistic Navier-Stokes expressions for the dissipative quantities, using Eq.~\eqref{NCL:Beq_RTA} we evaluate the non-equilibrium corrections to the phase-space distribution functions up to first-order in hydrodynamic gradients. In this way, for particles we get
\begin{eqnarray}\label{KT:delf1}
    \delta f_1 = &-& \frac{\tau_{\rm eq}}{\left(u\!\cdot\! p\right)} \!\left[ p^\alpha \partial_\alpha + \frac{m\, \chi}{2} \left( \partial^\alpha F^{\beta\gamma} \right) s_{\beta\gamma} \partial^{(p)}_\alpha \right]\! f_{\rm eq} \nn\\&+& \frac{\tau_{\rm eq}}{\left(u\!\cdot\! p\right)} \mathfrak{q} F^{\alpha\beta} p_\beta \partial^{(p)}_\alpha \!\left[ \frac{\tau_{\rm eq}}{\left(u\!\cdot\! p\right)} \!\left\{ p^\rho \partial_\rho + \frac{m\, \chi}{2} \left( \partial^\rho F^{\phi\kappa} \right) s_{\phi\kappa} \partial^{(p)}_\rho \right\}\! f_{\rm eq} \right], 
\end{eqnarray}
where, $\partial^{(p)}_\alpha\equiv \frac{\partial}{\partial p^\alpha}$ is the partial derivative with respect to particle momenta. Anti-particle analogue of $\delta f_1$ may be obtained from Eq.~(\ref{KT:delf1}) by the replacement $f\to \Bar{f}$, $\xi \to - \xi$, $\mathfrak{q}\to - \mathfrak{q}$ and, $\chi \to - \chi$. 

Substituting the non-equilibrium corrections to distribution functions in Eqs.~\eqref{diss1}-\eqref{diss4}, we get the following general form of  constitutive relations for the currents $X^{\mu_1 \dots \mu_s}\in \{n^{\mu}, \Pi, \pi^{\mu\nu}, \delta S^{\lambda,\mu\nu}\}$ at first order in gradients
\begin{align}\label{nmu_final}
    %1
    X^{\mu_1\dots\mu_s} = &\tau_{\mathrm{eq}} \left[ \beta_{X\Pi}^{\mu_1\dots\mu_s}\, \theta 
    + \beta_{Xa}^{{\mu_1\dots\mu_s}\alpha} \dot{u}_{\alpha}
    + \beta_{Xn}^{{\mu_1\dots\mu_s}\alpha} \left( \nabla_\alpha\xi \right) 
    + \beta_{XF}^{{\mu_1\dots\mu_s}\alpha\beta} \left( \nabla_{\alpha} B_{\beta} \right)\right.\\
    & \nn\left.~~~~~~~~~~~~~~~~~~~~~+ \beta_{X\pi}^{{\mu_1\dots\mu_s}\alpha\beta} \sigma_{\alpha\beta}
    + \beta_{X\Omega}^{{\mu_1\dots\mu_s}\alpha\beta} \Omega_{\alpha\beta} 
    + \beta_{X\Sigma}^{{\mu_1\dots\mu_s}\alpha\beta\gamma} \left( \nabla_\alpha \omega_{\beta\gamma} \right) \right], 
\end{align}
where we used the notation: $\theta\equiv \partial_\alpha u^\alpha$, $\dot{X}\equiv u^\alpha\partial_\alpha X$, $\nabla^\mu \equiv \partial^{\langle\mu\rangle}$, $\sigma^{\mu\nu}\equiv \partial^{\langle\mu} u^{\nu\rangle}$ and $\Omega_{\mu\nu} \equiv (\partial_\mu u_\nu - \partial_\nu u_\mu )/2$. 
The explicit expressions for the tensorial transport coefficients $\beta$ may be found in Ref.~\cite{Bhadury:2022ulr}. Here it is sufficient to note that the dissipative currents are affected by various hydrodynamic gradients, including those of magnetic field.
%
%%%%%%%%%%%%%%%%%%%%%%%%%%%%%%%%
\section{Discussion}
\label{sec:disc}
%%%%%%%%%%%%%%%%%%%%%%%%%%%%%%%%
%
Based on the above formalism we make some important observations and conclusions:
\begin{enumerate}
    \item \textit{\textbf{Relativistic Barnett and Einstein-de Haas effects.}}  
    Plugging equilibrium distribution functions into Eq.~\eqref{mag-pol-kin} one may show that the equilibrium magnetization tensor reads \cite{Bhadury:2022ulr}
    \begin{equation}
        \label{Meq}
        M^{\mu\nu}_{\rm eq}=a_1\,\omega^{\mu\nu} + a_2\, u^{[\mu} u_\gamma \omega^{\nu]\gamma}.
    \end{equation}
     Since in global equilibrium, the spin polarization tensor $\omega$ corresponds to the thermal vorticity tensor $\varpi$ \cite{Becattini:2007nd, Becattini:2009wh, Florkowski:2018fap, Florkowski:2017ruc, Florkowski:2018ahw, Florkowski:2019qdp, Florkowski:2021wvk, Florkowski:2017dyn}, from Eq.~(\ref{Meq}) we conclude that the vorticity of the fluid is related to its magnetization. Hence, Eq.~(\ref{Meq}) leads to relativistic analogs of the well-known Barnett \cite{Barnett} and  Einstein-de Haas \cite{Einstein} effects.
\item \textit{\textbf{Spin polarization due to the coupling between thermal vorticity and electromagnetic field.}} Using Eq.~\eqref{eq:LS}, one may derive the following evolution equation for $\omega^{\mu\nu}$
\begin{eqnarray}
    &\Dot{\omega}^{\mu\nu} \!=\! \mathcal{D}_{\Pi}^{[\mu\nu]}\, \theta 
    +\! \mathcal{D}_{a}^{[\mu\nu]\gamma} \Dot{u}_\gamma 
    +\! \mathcal{D}_{\mathrm{n}}^{[\mu\nu]\gamma} \left( \nabla_\gamma \xi \right) 
    +\! \mathcal{D}_{B}^{[\mu\nu]\rho\kappa}\! \left( \nabla_{\rho} B_{\kappa} \right)\label{omegadot}\\
    %2
    & \nonumber~~~~~~~~~~~~~~~~~~~~~~~~~~+ \mathcal{D}_{\pi}^{[\mu\nu]\rho\kappa} \sigma_{\rho\kappa} 
    + \mathcal{D}_{\Omega}^{[\mu\nu]\rho\kappa} \Omega_{\rho\kappa}\! 
    +\! \mathcal{D}_{\Sigma}^{[\mu\nu]\phi\rho\kappa}\! \left( \nabla_\phi \omega_{\rho\kappa} \right)\!, 
\end{eqnarray}
where the tensorial coefficients, $\mathcal{D}$, contain equilibrium quantities, see Ref.~\cite{Bhadury:2022ulr}. From Eq.~(\ref{omegadot}) we observe that among different gradient terms, there is a coupling of spin polarization tensor to the fluid vorticity represented by $\Omega$. The coefficient $\mathcal{D}_{\Omega}$ multiplying this term vanishes when the electromagnetic field is absent which implies that the conversion between spin polarization and vorticity proceeds via coupling with electromagnetic field.
\item \textit{\textbf{\textit{\textbf{Dissipative gradient terms.}}}} Demanding the positivity of the divergence of the entropy current (given by the Boltzmann H-theorem) one can show that only the following gradient terms in Eqs.~\eqref{nmu_final} are dissipative
\begin{eqnarray}
\Pi = - \zeta \theta, \qquad n^{\mu}\!\!&=&\!\!\kappa^{\mu\alpha} \left( \nabla_{\alpha} \xi \right), \qquad \pi^{\mu\nu} = \eta^{\mu\nu\alpha\beta} \sigma_{\alpha\beta}, \label{Smu_diss1}\\
\delta S^{\mu,\alpha\beta}\!\!&=&\!\!\Sigma^{\mu\alpha\beta\lambda\gamma\rho} \left(\nabla_\lambda \omega_{\gamma\rho} \right), \label{Smu_diss2}
\end{eqnarray}
where, comparing Eq.~\eqref{nmu_final} and Eqs.~\eqref{Smu_diss1}-\eqref{Smu_diss2}, the dissipative transport coefficients read: $\zeta=-\tau_{\mathrm{eq}} \beta_{\Pi\Pi}$, $\kappa^{\mu\alpha}=\tau_{\mathrm{eq}}\beta_{nn}^{\langle\mu\rangle\alpha}$, $\eta^{\mu\nu\alpha\beta}=\tau_{\mathrm{eq}}\beta_{\pi\pi}^{\langle\mu\nu\rangle\alpha\beta}$ and $\Sigma^{\lambda\mu\nu\alpha\beta\gamma}=\tau_{\mathrm{eq}}B_{\Sigma}^{\lambda,[\mu\nu]\alpha\beta\gamma}$.  
\end{enumerate}
%
%%%%%%%%%%%%%%%%%%%%%%%%%%%%%%%%
\section{Summary and outlook}
\label{sec:summary}
%%%%%%%%%%%%%%%%%%%%%%%%%%%%%%%%
%
In this work, we reviewed a recently developed framework of relativistic dissipative non-resistive magnetohydrodynamics for spin-polarized particles. Using the relativistic kinetic equation for the distribution function in an extended phase space of space-time position, momentum, and spin with the kinetic kernel treated in the relaxation time approximation, we calculated equations of motion for dissipative currents at first-order in gradients. The resulting equations of motion contain various transport coefficients, both dissipative and non-dissipative, which were distinguished using the positivity of the entropy production law. We have shown the emergence of the coupling between the magnetization and the vorticity of the fluid, which constitutes a mechanism leading to relativistic analogs of the Einstein-de Hass and Barnett effects. Furthermore, our analysis reveals that the relationship between magnetic fields and spin polarization occurs at the gradient level. In the context of relativistic heavy-ion collisions, our model offers a new perspective on explaining the splitting of the polarization signal for $\Lambda$ and anti-$\Lambda$ particles commonly attributed to the interaction between the magnetic field and the intrinsic magnetic moments of the emitted particles.

\smallskip
\noindent
{\it Acknowledgements:} A.J. was supported in part by the DST-INSPIRE faculty award under Grant No. DST/INSPIRE/04/2017/000038. This research was supported in part by the Polish National Science Centre Grants No. 2018/30/E/ST2/00432 (R.R.) and 2022/47/B/ST2/01372 (W.F.).
%%%%%
%%%%%
\bibliographystyle{utphys}
\bibliography{refs.bib}{}
\end{document}

%% file: commands.tex
% trigonometric functions   

% equation environments 
\def\be{\begin{equation}}
	\def\ee{\end{equation}}
	\def\barr{\begin{array}}
	\def\earr{\end{array}}
\def\beq{\begin{eqnarray}}
	\def\eeq{\end{eqnarray}}
\def\bfig{\begin{figure}}
	\def\efig{\end{figure}}
\newcommand{\bea}{\begin{eqnarray}}
	\newcommand{\eea}{\end{eqnarray}}

\def\LB{\left(}
\def\RB{\right)}
\def\LSB{\left[}
\def\RSB{\right]}
\def\LAB{\langle}
\def\RAB{\rangle}

\newcommand{\VP}{\vphantom{\frac{}{}}\!}
\def\lt{\left}
\def\rt{\right}
\def\CHI{\chi}
% \newcommand{\nn}{\nonumber}
% %%% nice and ordinary fractions
% \newcommand{\f}[2]{\frac{#1}{#2}}
% \newcommand{\onehalf}{{\nicefrac{1}{2}}}
% \newcommand{\onethird}{{\nicefrac{1}{3}}}
% \newcommand{\fivetwo}{{\nicefrac{5}{2}}}
%%% derivatives
% \newcommand{\p}{\partial}
% \newcommand{\del}{\partial}
%%% trace
%\newcommand{\tr}{{\rm tr}}
%%% references to equations
% \newcommand{\rf}[1]{Eq.~(\ref{#1})}
% \newcommand{\rfm}[1]{Eqs.~(\ref{#1})}
% \newcommand{\rftwo}[2]{Eqs.~(\ref{#1})~and~(\ref{#2})}
\newcommand{\rfmtwo}[2]{Eqs.~(\ref{#1})-(\ref{#2})}
\newcommand{\rfcs}[1]{Refs.~\cite{#1}}
% colors
% \newcommand{\red}{\color{red}}
% \newcommand{\blue}{\color{blue}}
% \newcommand{\green}{\color{green}}

% indices -- do not use for other purposes!
\def\a{\alpha}
\def\b{\beta}
\def\g{\gamma}
\def\d{\delta} 
\def\r{\rho}
\def\s{\sigma}
\def\c{\chi} 
 \def\lam{\lambda} 
% brackets
\def\LR{\left(} % round
\def\RR{\right)}
\def\LS{\left[} % square
\def\RS{\right]}
\def\LC{\left{} % curly
\def\RC{\right}}
\def\LA{\left\langle}% angle
\def\RA{\right\rangle}
\def\LD{\left.}% dummy
\def\RD{\right.}
% \newcommand{\dotb}[1]{\dot{\llbracket} #1 \rrbracket} % double
% phantoms
\def\HP{\hphantom{\alpha}} % horizontal

% trigonometric functions 

% \newcommand{\sh}[1]{\sinh#1}
% \newcommand{\ch}[1]{\cosh#1}
% \newcommand{\shb}[1]{\sinh\LR#1\RR}
% \newcommand{\chb}[1]{\cosh\LR#1\RR}
% \newcommand{\tU}{\theta_U}
% \newcommand{\tX}{\theta_X}
% \newcommand{\tY}{\theta_Y}
% \newcommand{\tZ}{\theta_Z}
% \newcommand{\tI}[1]{\theta_{#1}}
% \newcommand{\dU}{d_U}
% \newcommand{\dX}{d_X}
% \newcommand{\dY}{d_Y}
% \newcommand{\dZ}{d_Z}
% \newcommand{\dI}[1]{d_{#1}}

% fractions 
%\newcommand{\f}[2]{\frac{#1}{#2}}
\def\half{\frac{1}{2}}

% labels
\def\GLW{{\rm GLW}}
\def\LRFF{{\rm LRFF}}

% thermodynamic functions

% for unpolarized
\def\nU{n_{(0)}}
\def\nUi{n_{(0),i}}
\def\eU{\varepsilon_{(0)}}
\def\eUi{\varepsilon_{(0),i}}
\def\PU{P_{(0)}}
\def\PUi{P_{(0),i}}
\def\sU{s_{(0)}}
\def\sU{s_{(0),i}}

% for polarized
\def\nP{n_{}}
\def\eP{\varepsilon_{}}
\def\PP{P_{}}
\def\sP{s_{}}
\def\wP{w_{}}

% some commands
% \newcommand{\lab}[1]{\label{#1}}
\def\nn{\nonumber}

% references
% \newcommand{\refb}[1]{(\ref{#1})}
% \newcommand{\refeq}[1]{Eq.~(\ref{#1})}
% \newcommand{\refeqs}[1]{Eqs.~(\ref{#1})}

% calligraphic vars for thermodynamics

\def\cA{{\cal A}}
\def\cB{{\cal B}}
\def\cC{{\cal C}}
\def\cD{{\cal D}}
\def\cN{{\cal N}}
\def\cE{{\cal E}}
\def\cP{{\cal P}}
\def\cS{{\cal S}}
\def\cT{{\cal T}}
\def\cQ{{\cal Q}}
\def\cNN{{\cal N}_{(0)}}
\def\cEN{{\cal E}_{(0)}}
\def\cPN{{\cal P}_{(0)}}
\def\cSN{{\cal S}_{(0)}}

% three-vectors

% p four-vector
\def\pmu{p^\mu}
\def\pnu{p^\nu}

\def\vv{{\boldsymbol v}}
\def\pv{{\boldsymbol p}}
\def\av{{\boldsymbol a}}
\def\bv{{\boldsymbol b}}
\def\kv{{\boldsymbol k}}
%%%%%%%%%%% 
% omega tensor
\def\omnL{\omega_{\mu\nu}}
\def\omnU{\omega^{\mu\nu}}
\def\omnLbar{{\bar \omega}_{\mu\nu}}
\def\omnUbar{{\bar \omega}^{\mu\nu}}
\def\omnLbardot{{\dot {\bar \omega}}_{\mu\nu}}
\def\omnUbardot{{\dot {\bar \omega}}^{\mu\nu}}

\def\oabL{\omega_{\alpha\beta}}
\def\oabU{\omega^{\alpha\beta}}
\def\omnLD{{\tilde \omega}_{\mu\nu}}
\def\omnUD{\tilde {\omega}^{\mu\nu}}
\def\omnLDbar{{\bar {\tilde \omega}}_{\mu\nu}}
\def\omnUDbar{{\bar {\tilde {\omega}}}^{\mu\nu}}
\def\CHI{\chi}
\def\bchem{\mu_{\rm B}}
\def\bfug{\xi_{\rm B}}
\def\tfug{\xi}
%%%%%%%%%%%%%%%%%%%%
% New commands
%%%%%%%%%%%%%%%%%% 

% equation environments  
\def\be{\begin{equation}}
\def\ee{\end{equation}}
\def\ba{\begin{eqnarray}}
\def\ea{\end{eqnarray}}   

% indices -- do not use for other purposes!
\def\a{\alpha}
\def\b{\beta}
\def\g{\gamma}
\def\d{\delta} 
\def\r{\rho}
\def\s{\sigma}
\def\c{\chi}
 
% brackets
\def\LR{\left(} % round
\def\RR{\right)}
\def\LS{\left[} % square
\def\RS{\right]}
\def\LC{\left{} % curly
\def\RC{\right}}
\def\LA{\left\langle}% angle
\def\RA{\right\rangle}
\def\LD{\left.}% dummy
\def\RD{\right.}
% fractions 
%\newcommand{\f}[2]{\frac{#1}{#2}}
\def\half{\frac{1}{2}}

% labels
\def\GLW{{\rm GLW}}
\def\LRF{{\rm LRF}}

% thermodynamic functions

% for unpolarized
\def\nU{n_{(0)}}
\def\eU{\varepsilon_{(0)}}
\def\PU{P_{(0)}}
\def\sU{s_{(0)}}

% for polarized
\def\nP{n_{}}
\def\eP{\varepsilon_{}}
\def\PP{P_{}}
\def\sP{s_{}}
\def\wP{w_{}}

% three-vectors

% p four-vector
\def\pmu{p^\mu}
\def\pnu{p^\nu}

\def\vv{{\boldsymbol v}}
\def\pv{{\boldsymbol p}}
\def\av{{\boldsymbol a}}
\def\bv{{\boldsymbol b}}
\def\cv{{\boldsymbol c}}
\def\Cv{{\boldsymbol C}}
\def\kv{{\boldsymbol k}}
\def\piv{{\boldsymbol \pi}}

\def\thetap{\theta_\perp}
%%%%%%%%%%% 
% omega tensor
\def\omnL{\omega_{\mu\nu}}
\def\omnU{\omega^{\mu\nu}}
\def\omnLbar{{\bar \omega}_{\mu\nu}}
\def\omnUbar{{\bar \omega}^{\mu\nu}}
\def\omnLbardot{{\dot {\bar \omega}}_{\mu\nu}}
\def\omnUbardot{{\dot {\bar \omega}}^{\mu\nu}}

\def\oabL{\omega_{\alpha\beta}}
\def\oabU{\omega^{\alpha\beta}}
\def\omnLD{{\tilde \omega}_{\mu\nu}}
\def\omnUD{\tilde {\omega}^{\mu\nu}}
\def\omnLDbar{{\bar {\tilde \omega}}_{\mu\nu}}
\def\omnUDbar{{\bar {\tilde {\omega}}}^{\mu\nu}}

% Levi-Civita tensor
\def\epsLmnbg{\epsilon_{\mu\nu\beta\gamma}}
\def\epsUmnbg{\epsilon^{\mu\nu\beta\gamma}}
\def\epsLmnab{\epsilon_{\mu\nu\alpha\beta}}
\def\epsUmnab{\epsilon^{\mu\nu\alpha\beta}}

\def\epsUmnrs{\epsilon^{\mu\nu\rho \sigma}}
\def\epsUlnrs{\epsilon^{\lambda \nu\rho \sigma}}
\def\epsUlmrs{\epsilon^{\lambda \mu\rho \sigma}}

\def\epsLmnbg{\epsilon_{\mu\nu\beta\gamma}}
\def\epsUmnbg{\epsilon^{\mu\nu\beta\gamma}}
\def\epsLmnab{\epsilon_{\mu\nu\alpha\beta}}
\def\epsUmnab{\epsilon^{\mu\nu\alpha\beta}}

\def\epsLabgd{\epsilon_{\alpha\beta\gamma\delta}}
\def\epsUabgd{\epsilon^{\alpha\beta\gamma\delta}}

\def\epsUmnrs{\epsilon^{\mu\nu\rho \sigma}}
\def\epsUlnrs{\epsilon^{\lambda \nu\rho \sigma}}
\def\epsUlmrs{\epsilon^{\lambda \mu\rho \sigma}}

\def\epsLijk{\epsilon_{ijk}}
%%%%%%%%%%%%%

%%%%%%%%%%%%%%%%%%%%%%%%%%%%%%%%%%%%%%%%%%%%%%%%%%%%%%%%%%%%%%%%%%%%%%%%%%%%%%%%%%%%%%%%%%%%%%%%
% custom commands
%%%%%%%%%%%%%%%%%%%%%%%%%%%%%%%%%%%%%%%%%%%%%%%%%%%%%%%%%%%%%%%%%%%%%%%%%%%%%%%%%%%%%%%%%%%%%%%%

% Levi-Civita tensor
\def\epsLmnbg{\epsilon_{\mu\nu\beta\gamma}}
\def\epsUmnbg{\epsilon^{\mu\nu\beta\gamma}}
\def\epsLmnab{\epsilon_{\mu\nu\alpha\beta}}
\def\epsUmnab{\epsilon^{\mu\nu\alpha\beta}}

\def\epsUmnrs{\epsilon^{\mu\nu\rho \sigma}}
\def\epsUlnrs{\epsilon^{\lambda \nu\rho \sigma}}
\def\epsUlmrs{\epsilon^{\lambda \mu\rho \sigma}}

\def\epsLmnbg{\epsilon_{\mu\nu\beta\gamma}}
\def\epsUmnbg{\epsilon^{\mu\nu\beta\gamma}}
\def\epsLmnab{\epsilon_{\mu\nu\alpha\beta}}
\def\epsUmnab{\epsilon^{\mu\nu\alpha\beta}}

\def\epsLabgd{\epsilon_{\alpha\beta\gamma\delta}}
\def\epsUabgd{\epsilon^{\alpha\beta\gamma\delta}}

\def\epsUmnrs{\epsilon^{\mu\nu\rho \sigma}}
\def\epsUlnrs{\epsilon^{\lambda \nu\rho \sigma}}
\def\epsUlmrs{\epsilon^{\lambda \mu\rho \sigma}}

\def\epsLijk{\epsilon_{ijk}}
% fractions 
\def\half{\frac{1}{2}}
% labels
\def\GLW{{\rm GLW}}

% thermodynamic functions
\def\n0{n_{(0)}}
\def\e0{\varepsilon_{(0)}}
\def\P0{P_{(0)}}
\newcommand{\redflag}[1]{{\color{red} #1}}
\newcommand{\blueflag}[1]{{\color{blue} #1}}
\newcommand{\checked}[1]{{\color{darkblue} \bf [#1]}}
\newcommand{\Psis}{{\sf \Psi}}
\newcommand{\psis}{{\sf \psi}}
\newcommand{\Psibar}{{\overline \Psi}}
%% NAMES %%
\def\eMf{electromagnetic (EM) }
\def\EMf{Electromagnetic (EM) }
\def\EM{EM }
\def\lRFf{local rest frame (LRF)}
\def\LRFf{Local rest frame (LRF) }
\def\LRF{LRF }
\def\QGPf{Quark gluon plasma (QGP) }
\def\qGPf{Quark gluon plasma (QGP) }
\def\QGP{QGP }
\def\mHDf{magnetohydrodynamic (MHD) }
\def\MHDf{Magnetohydrodynamic (MHD) }
\def\MHD{MHD }
\def\iMHD{iMHD }
\def\HD{Hydrodynamics }
\def\hD{hydrodynamics }
\def\RHD{Relativistic hydrodynamics }
\def\rHD{relativistic hydrodynamics }
\def\rMHDf{relativistic magnetohydrodynamic (RMHD) }
\def\RMHDf{Relativistic magnetohydrodynamic (RMHD) }
\def\RMHD{RMHD }
\def\eOMf{equations of motion (EOM)~}
\def\EOMf{Equations of motion (EOM)~}
\def\EOM{EOM}
\def\fl{\ensuremath{\text{Fluid}}}
\def\lrf{\ensuremath{\text{LRF}}}
\def\BVf{Boltzmann-Vlasov (BV) }
\def\BV{BV\,}
%%%%%%%%%%%%%%%%%%%%%%%%%%%%%%%%%%%%%%%%%%
% DISTRIBUTION FUNCTIONS OF VARIOUS KINDS
%%%%%%%%%%%%%%%%%%%%%%%%%%%%%%%%%%%%%%%%%%
		
\def\rhoLEQ{{\widehat{\rho}}_{\rm \small LEQ}}
\def\rhoGEQ{{\widehat{\rho}}_{\rm \small GEQ}}
		
\def\fplushat{{\hat f}^+}
\def\fminushat{{\hat f}^-}
		
\def\fplusrs{f^+_{rs}}
\def\fplussr{f^+_{sr}}
\def\fplusrsxp{f^+_{rs}(x,p)}
\def\fplussrxp{f^+_{sr}(x,p)}
		
\def\fminusrs{f^-_{rs}}
\def\fminussr{f^-_{sr}}
\def\fminusrsxp{f^-_{rs}(x,p)}
\def\fminussrxp{f^-_{sr}(x,p)}
		
\def\fpmrs{f^\pm_{rs}}
\def\fpmrsxp{f^\pm_{rs}(x,p)}

\def\feqplus{f^+_{eq}}
\def\feqplus{f^+_{eq}}
\def\feqplusxp{f^+_{eq}(x,p)}
\def\feqplusxp{f^+_{eq}(x,p)}
		
\def\feqminus{f^-_{eq}}
\def\feqminus{f^-_{eq}}
\def\feqminusxp{f^-_{eq}(x,p)}
\def\feqminusxp{f^-_{eq}(x,p)}
	
\def\feqpm{f^\pm_{{\rm eq}}}
\def\feqpmxp{f^\pm_{{\rm eq}}(x,p)}
\def\feqpmi{f^\pm_{{\rm eq},i}}
\def\feqpmxpi{f^\pm_{{\rm eq},i}(x,p)}
\def\fpm{f^\pm}
\def\fpmxp{f^\pm(x,p)}
\def\fpmi{f^\pm_i}
\def\fpmxpi{f^\pm_i(x,p)}
\newcommand{\rs}[1]{\textcolor{red}{#1}}
\newcommand{\rrin}[1]{\textcolor{blue}{#1}}
\newcommand{\rrout}[1]{\textcolor{blue}{\sout{#1}}}
\newcommand{\lie}[2]{\pounds_{#1}\,#2}
\newcommand{\rd}{\mathrm{d}}
\def\re{\mathrm{e}}
\def\echarge{\ensuremath{\rho_e}}
\def\cond{\ensuremath{{\sigma_e}}}
\newcommand{\msnote}[1]{\todo[author=Masoud]{#1}}
\newcommand{\msnotei}[1]{\todo[author=Masoud,inline]{#1}}
\newcommand{\explainindetail}[1]{\todo[color=red!40]{#1}}
\newcommand{\insertref}[1]{\todo[color=green!40]{#1}}
\newcommand{\fm}{\rm{\,fm}}
\newcommand{\fmc}{\rm{\,fm/c}}
%%%new commands

% u four-vector

\def\uv{{\boldsymbol U}}

% k four-vector

\def\kbarzero{ {\bar k}^0}
\def\kv{{\boldsymbol k}}
\def\kbarv{{\bar {\boldsymbol k}}}
% omega four-vector

\def\obarzero{ {\bar \omega}^0}
\def\ov{{\boldsymbol \omega}}
\def\obar{{\bar \omega}}
\def\obarv{{\bar {\boldsymbol \omega}}}
% e and b three-vectors

\def\ev{{\boldsymbol e}}
\def\bv{{\boldsymbol b}}
%%%% For appendix
\newcommand{\tT}{\theta_T}
\newcommand{\UD}[1]{\oU{#1}}
\newcommand{\XD}[1]{\oX{#1}}
\newcommand{\YD}[1]{\oY{#1}}
\newcommand{\ZD}[1]{\oZ{#1}}
\newcommand\oU[1]{\ensurestackMath{\stackon[1pt]{#1}{\mkern2mu\bullet}}}
\newcommand\oX[1]{\ensurestackMath{\stackon[1pt]{#1}{\mkern2mu\star}}}
\newcommand\oY[1]{\ensurestackMath{\stackon[1pt]{#1}{\mkern2mu\smwhitestar}}}
\newcommand\oZ[1]{\ensurestackMath{\stackon[1pt]{#1}{\mkern2mu\circ}}}
% other
\def\Aone{{ \cal A}_1 }
\def\Atwo{{ \cal A}_2 }
\def\Athree{{ \cal A}_3 }
\def\Afour{{ \cal A}_4 }
\def\vv{{\boldsymbol v}}
\def\pv{{\boldsymbol p}}

\newcommand{\inv}[1]{\frac{1}{#1}}
\newcommand{\iinv}[1]{1/#1}